\documentstyle[aps,twocolumn]{revtex}
\input{epsf}

\def\be{\begin{eqnarray}}
\def\ee{\end{eqnarray}}
\def\a{\alpha}
\def\b{\beta}

\def\l{\lambda}
\def\o{\omega}
\def\s{\sigma}
\def\o{\omega}
\def\ol{\overline}
\def\e{\epsilon}
\def\f{\varphi}
\def\t{\theta}
\def\r{\rho}
\def\p{\partial}

\begin{document}
\title{Magnetophoresis  of neutral particles in AC electric field.}
\author{ A.A.~Ovchinnikov }
\address{\it  Max Planck Institute for Physics of Complex Systems, Dresden.\\
Joint Institute of Chemical Physics, Moscow.}
\wideabs{
\maketitle

\begin{abstract}

It has been shown that under joint action of DC magnetic and AC
electric fields the neutral particles (atoms, molecules, nano-size particles
etc.)  nearby interface move with permanent velocity along the surface.
The physical reason for such transport is due to an 
essentially non-linear dynamics of particles on the surface. The driving force
depends on non-linear dynamical properties of polarisation
of the particles and a coupling between the polarisation  and the position
of the particles on the surface. An evaluation of the particle velocity as
function of the external AC electric and magnetic  fields is given. This
effect leads to an appearance of a DC neutral current in a layer of dielectric
liquid deposited on metallic surface in the permanent magnetic and the AC
electric fields. The conditions for an observation of this current are
discussed. 

\end{abstract}
}


The directed transport of particles under action of AC driving force is now
well established phenomenon in various non-linear systems. Most detailed this
transport has been studied for one-dimensional motion of particle in the
space-periodic potential $U(x)$ and a time-periodic force having general form
$F(t)=F_0 \cos{\o t}+F_1 \cos{(2\o t+\t)}$. (See review \cite{1} and recent
works \cite{2,3,4,5}.) Main physical reason for the directed motion of the
particles is connected with the so called latent asymmetry of driving force
$F(t)$. Though the time average of $F(t)$ is zero the time average
$\overline {F^3(t)}$ doesn't vanish and depends on $\t$. Non-linearity of
the system with potential $U(x)$ leads to mixing of harmonics such a way that
average velocity is not zero and directed along $\overline {F^3(t)}$.

The similar phenomenon which we call non-linear Hall effect has been described
in \cite{6}. It turns out that in metals and semiconductors the joint action
of permanent magnetic and an AC electric fields leads to arising of DC surface
electric current. The physical reason for such a current is due  to
essentially non-linear dynamics of electronic liquid in the surface layer. 
Consider dynamics of electronic gas in skin layer of metal under
an influence of external AC electric field $E_0(t)=E_0\cos{\o t}$
directed perpendicularly to the surface of the
metal (along z axis) and in the magnetic field $H$. The electronic gas vibrates
perpendicularly to the surface. Due to the nonlinearity of these vibrations a
normal to the surface velocity $v(z,t)$ and density of electron gas $\r (z,t)$
are periodic functions of time with frequency $\o$ having all overtones:
$$
v=v_1(z)\cos{(\o t+\f_1)}+v_2(z)\cos{(2\o t+\f_2)}\ldots
$$
$$
\r=\r_0+\r_1(z)\cos{(\o t+\f_3)}+\r_2(z)\cos{(2\o t+\f_4)}\ldots
$$
The above amplitudes and phases of harmonics as functions of $z$  are a
subject of calculations. 
The Lorentz forces induced by magnetic field lead to tangential vibration of
electronic gas with amplitude which is proportional to this magnetic field. 
Corresponding tangential velocity $u(z,t)$ has the same form as $v(z,t)$ with
its own amplitudes and phases. The density of tangential current is
$j(z,t)=u(z,t)\r(z,t)$. The time averaging of $j(z,t)$ over period gives the
DC component of the current. 

The aim of this paper is to show that the analogous  neutral current arises if
a neutral (instead of a charged liquid ) dielectric particle deposited on
metallic surface. 

Let us consider an adatom or a molecule on the metallic  surface under action
of the periodic electric field $E_0(t)=E_0\cos{\o t}$ directed perpendicularly
to the surface. An induced $z$-component of the dipole moment $d(t)$ of the
particle is also a periodic 
function of time. Due to an anharmonicity the $d(t)$ has all possible
overtones and generally can be written as 
\be
\begin{array}{ll}
d(t)&=d_{0}+d_{1}\cos{(\o t+\f_1)}+\\
&d_{2}\cos{(2\o t+\f_2)}+\ldots\\
\end{array}
\label{1}
\ee
where $d_0$ being a dipole moment of the particle adsorbed on the surface and
$d_1 , d_2 , d_3$ are the induced dipole moments of different order with
respect to the external electric field $E_0(t)$. The phases $\f_i$ are defined
by an inner dynamics of the particle. In a magnetic field H directed along the
surface (axis $y$ ) the tangential Lorentz force $F_L(t)$ acting on the
particle can be expressed as
\be
F_L(t)={1 \over c}H{\dot d} 
\label{2}
\ee
Here $c$ is a light velocity. Then the Newton-Langeveen equation for tangent
velocity $v$ of the particle could be written as following 
\be
m {{dv} \over dt}=F_L(t)-{\p U \over \p x}-
\nu v+f_r(t)
\label{3}
\ee
where   $m$ is a total mass of the particle, $U(x)$ is a space periodic
potential which could be chosen in simplest form $U(x)=U_0\sin{(x/a)}$ ($a$
is 
a lattice constant). Last term of r.s. of Eq.(3) is a damping originated either
from a finite conductivity of the metal or from the Stokes friction of the
particle if the latter moves in a liquid. The function $f_r(t)$  in (3) is a
random force with a spectrum of a white noise  included into (3) in order to
give a Maxwell distribution in a thermal equilibrium. 

The Equation (3) has been intensively  studied last decade in both classical
and quantum regimes \cite{1,2,3,4,5}. In over-damping regime Eq.(3) can be
analysed 
analytically. In this case Eq.(3) is equivalent to the diffusion equation for
a density of probability $w(x,t)$
\be
{{\p w(x,t)} \over {\p t}}=D{{\p} \over {\p x}}
\Bigl[{{\p w} \over {\p x}}+{1\over T}\Bigl(F_L(t)+
{{\p U} \over {\p x}}\Bigr)w\Bigr]
\label{4}
\ee
where $T$ is a temperature of the system and $D=T/\nu$. In reality
$F_L(t)a/T\ll 1$ (weak external electric  and magnetic fields). Density of
current $j(x,t)$ is equal to
$$ 
j(x,t)=-D\Bigl( {{\p w} \over {\p x}}+{1\over T}
\bigl(F_L(t)+{{\p U} \over {\p x}}\bigr)w\Bigr)
$$
and the particle velocity averaged over a period $T_0$  is 
$$ 
v={1\over T_0}\int_{-\infty}^{\infty}\int_0^{T_0}j(x,t)dxdt
$$
Usually the lattice potential $U(x)$ is also very small, $U_0/T\ll1$. In a
steady state regime $w(x,t)$ is a periodic function of time
with a frequency $\o$ and a space periodic function with period $a$. So, the
the $w(x,t)$ can be expressed via a Fourier set as
\be
w(x,t)=\sum_{n=-\infty}^{\infty} w_n(t)e^{inx/a}
\label{5}
\ee
where $w_0=1/a$ by definition and 
$$w_n(t)=w_{-n}^* (t)$$. 
At very small $F_L$ and $U_0$ one can neglect all components
$w_n(t)$ except $w_0$, $w_{\pm 1}$. Then the equation for $w_1(t)$ has the
following form
$$
{{dw_1}\over {dt}}=-{D\over {a^2}} w_1+i{F_L (t)D\over {aT}}w_1  +
i{U_0D \over {2a^2T}}
$$
The periodic solution of this equation can be readily found in a form 
\be
w_1(t)={{U_0D}\over {2a^3 T}}i\int_0^{\infty} \exp\Bigl\{-{\xi D \over a^2}
+iQ(\xi,t)\Bigr\}d\xi 
\label{6}
\ee
$$
Q(\xi,t)={{e^2DH}\over{a^2Tc}}[d(t)-d(t-\xi )]
$$
where function $d(t)$ has to be taken from (1). Making use an expansion of
the exponent of integrand up to the third order with respect to  magnetic
field $H$ one 
arrives to the following expression for the average velocity of particle
\be
v={{U_0^2D^2}\over {4a^3T^2}}\Biggl( {DH\over {Tac}}\Biggr)^3
{ d_1^2d_2\o \over {(D/a^2)^2+4\o^2}}\sin{\f_2(\o )}
\label{7}
\ee
For quantitative estimation of $v$ one has to find $d_1(\o), ~d_2(\o)$ and a
phase $\f_2(\o)$ considering nonlinear dynamics of induced dipole moment  of
particle 
deposited on metallic surface in external AC electric field $E_0(t)$. This
problem has been studied in all details in \cite{7} and here we give only the
results of this consideration. The simplest non-linear equation for the normal
component of the dipole moment $d(t)$ has the following view
\be
\begin{array}{ll}
{\ddot d} (t)+&\o^2_0 d(t)+k_3 d^2(t)+\\
&({e^2E_0/\mu})\cos{\o t}-\nu {\dot d}(t)=0\\
\end{array}
\label{8}
\ee
where $\o_0$ is an eigenfrequency of the dipole oscillator, $\mu$ is its mass,
$k_3$ is a coefficient connected with third order polarisability of the
particle and $\nu$ is an inverse relaxation time. The latter can be taken from
the Raman spectroscopy data for this particle on the surface of the metal
\cite{7}. Roughly it can be estimated as $\nu= e^2/(a^3\s_0\mu)$. Here
$\s_0$ is a conductivity of the metal. The simple analyses of the (8) at
small frequency of the external field ($\o\ll \o_0$) gives
$$
d_1={{E_0e^2}\over {\mu\o_0^2}},~~
d_2={{d_1^2 k_3}\over {\o_0^2}},~~
\f_2(\o)={{\nu \o}\over {\o_0^2}}\ll 1,~~
$$
Thus, the mean velocity of the particle $v$ is proportional to the $H^3$ and
$E^4$. Choosing the reasonable parameters for all physical values in (7) one
can see that estimated mean velocity $v$ is too small to give an observable
effect.

However, there exists another mechanism of an appearance of mean velocity which
gives $v$ linearly depending on the magnetic field $H$. 

First of all consider the dynamics of the dipole particle on the surface of
metal. The potential function $U(z,x,d)$ of the particle depends on three
coordinates $z,x$ and $d$. Here $z$ is a distance between centre of mass of
the particle and the surface of the metal; $x$ is a coordinate of the particle
along the surface and $d$ is a $z$ component of the dipole moment of the
particle. The $U(z,x,d)$ can be written as a sum of four components: $U_1(z)$,
$U_2(z,d)$, $U_3(d)$ and $U_4(z,x,d)$
\be 
U=\sum_{i=1}^4 U_i
\label{9}
\ee
where $U_1(z)$ includes a Van der Waals attraction and an exchange repulsion
forces between the surface and the physically adsorbed particles. The $
U_2(d,z)$ is a dipole-metal interaction which can be taken as dipole-dipole
interaction of the  particle dipole and its mirror image in a metal. Thus, we
have approximately
\be 
U_2(z,d)=-{d^2 \over {8z^3\e}}
\label{10}
\ee
where $\e$ is a dielectric constant. The $U_3(d)$ is a potential function
determining an inner dynamics of the particle dipole. And finally, the
$U_4(z,x,d)$ is a potential function which couples the normal (to the surface)
motion with a tangent coordinate $x$ of particle. If we neglect an atomic
structure of a surface this function also can be neglected. So, further for
simplicity we choose $U_4(z,x,d)=0$. The quantitative descriptions of all
these functions are given in many textbooks and monographs (see \cite{7}). For
our aims we limit ourself by an expansion of $U(z,x,d)$ in vicinity of the
equilibrium positions $z_{eq}$ and equilibrium dipole moment $d_{eq}$. Further
$z$ and $d$ denote deviations of these coordinates from their equilibrium
values. 

In this approximation the $U(z,x,d)$ can be rewritten as following 
\be
\begin{array}{ll}
U(z,x,d)={{\mu d^2\o^2_0}/{2e^2}}+ {{m z^2\o^2_p}/{2}}+
z d \b +&\\
~~~~~~~~\l_1d^3+\l_2 d^3+\l _3d^2 z+\l_3 dz^2+\l_4z^3&\\
\end{array}
\label{11}
\ee
Here $\o_0$ and $\o_p$ are corresponding eigenfrequencies and $\mu$ and $m$
are the mass of the dipole oscillator and the mass of the particle,
respectively.  

Due to the large difference between $\mu$ and $m$ the anharmonic terms are 
small and can be treated perturbatively.  The Newton equations of motion of the
particle in external magnetic and AC electric fields have the following form
\be
\begin{array}{ll}
m{\ddot z}=&-m\o_p^2 z-\b d-\l_2 d^2-\\
&2\l_3dz+3\l_4z^2 -\nu (z,d){\dot z}\\
\end{array}
\label{12}
\ee
\be
\begin{array}{ll}
\mu{\ddot d }=&-\mu\o_0^2 d-\b z -3\l_1 d^2 -2\l_2 dz-\\
&\l_3z^2 -\nu (z,d){\dot d}+e^2E_0\cos{\o t}\\
\end{array}
\label{13}
\ee
\be
m{\ddot x }={{{\dot d}H}\over c}-\nu (z,d){\dot x}
\label{14}
\ee
The first term in r.s. of (14) is the Lorentz force discussed above. Generally,
the damping coefficient $\nu (z,d)$ is a function of both dipole moment of the
particle $d$ and the position $z$ of the particle with respect to the surface
of the metal. The motion of the dipole on the vacuum - metal interface induces
the electric current inside of the metal. The density of this current depends
also on $d$ and $z$. A resistivity to this current is just the mechanism of
the damping. The detailed calculation of the function $\nu (d,z)$ is a quite
complicated electrodynamic  problem discussed in \cite{7}. However, for
qualitative estimation of  
it we can use a rough dimension consideration. Taking into account that 
$\nu (d,z)$ proportional to specific resistivity of the metal we arrive to the
following expression 
\be 
\nu (d,z)\sim {{e^2}\over {z_{eq}^3\s_0}}
\Bigl(1+{z\over z_{eq}}\a_1+{d\over d_{eq}}\a_2 +\ldots \Bigr)
\label{15}
\ee
where $\s_0(z)$ is a conductivity of metal. The $\a_1$, $\a_2$
are coefficients of an expansion of $\nu (d,z)$ with respect to $z$ and $d$
which we 
choose of order of unit. Below we shall see that most important for the effect
is coefficient $\a_1$, a sign of which determines the sign of the mean
velocity of the particle. As a rule, $\a_1,)$. 

At solution of equations (12-14) we use two approximations. First of all, the
anharmonic terms  were treated perturbatively. The second, we consider the
frequency of the external electric field $\o$ much less than the
eigenfrequency of the dipole oscillator ($\o\ll\o_0$). However, the $\o$ can be
compared with $\o_p$, the eigenfrequency of vibration of the adsorbed particle
on the metal. 

Thus, the equation (13) can be solved adiabatically. Ommiting the simple
perturbative calculations we get
\be
{\ol v} =-{{\a^2HE_0^2 \o^2\a_1 \b} \over
{2cz_{eq}[m^2(\o_p^2-\o^2)^2+ \nu _0^2\o^2]}}
\label{16}
\ee
Here $\a$ is a linear polarisability of the particle which is determined 
from a quantum calculation. 
Similarly to the non-linear Hall current \cite{6} the mean velocity ${\ol v}$
depends linearly on the magnetic field $H$ and quadratically on an amplitude
of external AC electric field. The frequency dependence of ${\ol v}$ 
has the resonance character reaching the maximum at $\o=\o_p$, the
eigenfrequency of an oscillation of the particle adsorbed on metal surface. At
very small $\o$ the effect trends to zero quadratically.
 So, the effect is quite similar to the nonlinear Hall effect described in
\cite{6}, but the current of the particle is the neutral one.

At resonance frequency in magnetic field of order of $\sim 1000$ Gauss and
electric field of order of $\sim 1000~ V/cm$ a value ${\ol v}$ can reach a few
centimetre per second. At this estimation we took for conductivity of metal
the value of order of $10^4~{S/cm}$. For more exact estimation
one needs to define quantitatively all parameters of the system adsorbed
particle- surface of the metal, which is quite complicated task having in mind
our limited knowledge of the system.

This consideration can be readily expanded to the case of the nano-size
dielectric particles deposited on metal surface. The size of the particle
comes into the main formula (16) the following way. First of all note that the
polarisability and mass of the particle are proportional to their volume. The
$\o_p^2$ is diminishing as $1/m$. The $\b$ depends on the radius of the
particle $R$ as $\b \sim 1/R^4$. Taking into account all these dependences we
can conclude that at resonance frequency $\o=\o_p$ the ${\ol v}$ will increase
quadratically with the the increasing of the particle radius. 

If instead of the selected particle adsorbed on the surface we consider the
layer of a dielectric fluid on the surface the described effect leads to an 
appearance of a surface current $I$ the value of which can be estimated as 
\be
I ={\ol v} \rho l
\label{17}
\ee
where $\rho$ is a density of the liquid, and $l$ is a width of the
surface layer which typically is about 100 $A$ for dielectric liquid like a
water.

One can give two suggestions for observation of the effect. First of all, the
Brownian motion of nano-size particles adsorbed on a 
transient metallic surface in permanent magnetic and AC electric fields has to
be a directed one. So, it will be interesting to repeat Perrin-like experiment
in these conditions. Secondly, if we take a cylindrical condenser  with a
fluid dielectric media in a magnetic field directed along the main axis of the
cylinder then the tangent flow of fluid could be observable. The intriguing
possibility arises if we choose as dielectric medium a superfluid
liquid. Corresponding estimations for ${\ol v}$ will be published elsewhere.

The main obstacle for observation of this effect 
seems to be connected with an absorption of ultrahigh frequency
electromagnetic field into material and its warming up. 
On the other hand, at low (or ultra-low) temperature the
absorption and warming up in AC electric field could be made small enough for
such observation.

Author thanks RFFI Grant N 00-15-97334 for support,  M.Ya.Ovchinnikova for
help in numerical calculations and 
S.Flach for numerous discussions on related problems.

\vspace{0.2in}


\vspace{0.2in}


\end{document}